\documentclass[prl, twocolumn, showpacs, preprintnumbers, amsmath, amssymb, superscriptaddress]{revtex4-1}
\usepackage{graphicx}           
\usepackage{dcolumn}            
\usepackage{bm}                 
\usepackage[bookmarks, colorlinks, linkcolor=blue, citecolor=blue, urlcolor=blue]{hyperref}

\begin{document}

\title{Three-dimensional atom localization by laser fields in a four-level tripod system}

\author{Vladimir S. Ivanov}
\email{ivvl82@gmail.com}
\affiliation{Turku Centre for Quantum Physics, Department of Physics and Astronomy, University of Turku, 20014 Turku, Finland}

\author{Yuri V. Rozhdestvensky}
\email{rozd-yu@mail.ru}
\affiliation{Saint Petersburg National Research University of Information Technologies, Mechanics and Optics, 197101 St. Petersburg, Russia}

\author{Kalle-Antti Suominen}
\email{Kalle-Antti.Suominen@utu.fi}
\affiliation{Turku Centre for Quantum Physics, Department of Physics and Astronomy, University of Turku, 20014 Turku, Finland}

\date{\today}

\begin{abstract}
  We present a scheme of the high-precise three-dimensional (3D) localization by the measurement of the atomic-level population. The scheme is applied to a four-level tripod-type atom coupled by three strong standing waves and a probe running wave. As a result, the atom can be localized in volumes that are substantially smaller than a cubic optical wavelength. The upper-level distribution depends crucially on the atom-field coupling and it forms 3D periodic structures composed of spheres, hourglasses, bowls, donuts, or deformed barrels.
\end{abstract}


\maketitle

The concept of Heisenberg microscope~\cite{Heisenberg1949} demonstrates that atoms can not be localized using light within the optical half-wavelength unless one takes advantage of the internal structure of atoms. One approach is to obtain spatially changing structures in level populations in a cloud of atoms. Motivated e.g.~by possible applications in atomic nanolithography~\cite{Agarwal2006}, different methods for going beyond this limit have been proposed over the years, including the measurement of the atomic resonance frequency~\cite{Thomas1990,Stokes1991,Gardner1993}, the phase shift~\cite{Storey1992,Storey1993,Storey1994}, or the atomic dipole~\cite{Kunze1994}. The highest theoretical localization degrees have been demonstrated in schemes that rely on measuring either spontaneous emission~\cite{Herkommer1997,Qamar2000a,Qamar2000b} or a level population~\cite{Paspalakis2001,Paspalakis2005} of an atom moving through a resonant standing wave.

In general, as a proof-of-principle, one-dimensional systems have been very popular, see e.g. the recent studies~\cite{Pancha2012,Wang2011262,Rahmatullah20131587,Dutta20131890,Dutta2013,Xu2013}. Studies of atomic systems in two dimensions have also been numerous recently~\cite{Wang20121132,Ding2011,Wan2011985,Li20113978,Wang2014263,Wan:11,PhysRevA.87.043816,PhysRevA.88.013846,Rahmatullah2014684,Ding:12,PhysRevA.83.063834,PhysRevA.84.043840}. It is important to note that in 2D, one can obtain, in addition to mere subwavelength localization, further spatial structuring of the atomic locations, see e.g.~\cite{Ivanov2010}. A key question is whether localization and structuring can be obtained also in 3D, and what kind of structures can appear. Possible applications may include high-precision position-dependent chemistry without a real change in the atomic distribution, combining state-selective localization with state-selective chemical reactions. In addition, 3D structuring produces a wide variety of practical realizations of such chemical reactions, demonstrating the necessity of reliable methods for the prediction of localization structures. For any realistic scheme one needs to carefully consider the atomic structure and to include spontaneous emission. Previously 3D localization has been discussed in a five-level system without taking the geometry of the atomic dipole coupling into account~\cite{Qi2012}. We show that 3D localization is realistically possible with only four atomic levels in a tripod scheme, and that one obtains interesting structures around points of localization by simple tuning of the parameters of the localizing fields.

We consider the $^{87}$Rb D$_2$ line (the $5^2\mathbf{S}_{1/2} \to 5^2\mathbf{P}_{3/2}$ transition), which is only thought as a possible example of tripod system with three ground states $| F = 1, m_F \rangle$ ($m_F = -1, 0, 1$) and upper state $| F' = 0, m'_F = 0 \rangle$. The Wigner-Eckart theorem~\cite{Brink1993} concerns that the covariant component $q$ of the dipole matrix is proportional to the Clebsch-Gordan coefficient $\langle F m_F | F' 1 m'_F q \rangle$, i.e.,
\begin{align}
  \langle F m_F | d_q | F' m'_F \rangle = \langle F \| \mathbf{d} \| F' \rangle \langle F m_F | F' 1 m'_F q \rangle.
\end{align}
The proportionality factor is factorized into the reduced matrix element $\langle J \| \mathbf{d} \| J' \rangle$ and a Wigner 6-$j$ symbol:
\begin{multline}
  \langle F \| \mathbf{d} \| F' \rangle = \langle J \| \mathbf{d} \| J' \rangle (-1)^{F'+J+1+I}
  \\
  \times \sqrt{(2F' + 1) (2J + 1)}
  \begin{Bmatrix}
    J & J' & 1
    \\
    F' & F & I
  \end{Bmatrix},
\end{multline}
where the $^{87}$Rb D$_2$ line is defined by the total angular momentum $I=3/2$ and the reduced matrix element
\begin{align}
  \langle J = 1/2 \| \mathbf{d} \| J' = 3/2 \rangle = 4.227(5)\mbox{ a.e.}
\end{align}
The Clebsch-Gordan coefficient is given by a 3-$j$ symbol:
\begin{align}
  \label{eq:clebsch-gordan}
  \langle F m_F | F' 1 m'_F q \rangle = (-1)^{F'-1+m_F} \sqrt{2F + 1}
  \begin{pmatrix}
    F' & 1 & F
    \\
    m'_F & q & -m_F
  \end{pmatrix}.
\end{align}

\begin{figure}
  (a)\\ \includegraphics{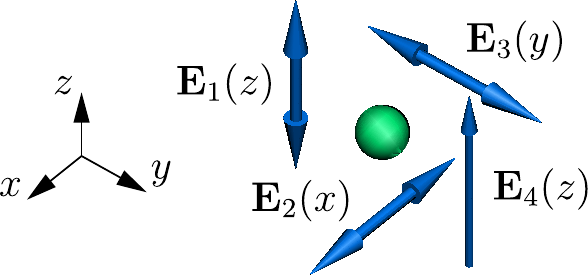}
  \\
  (b)\\ \includegraphics{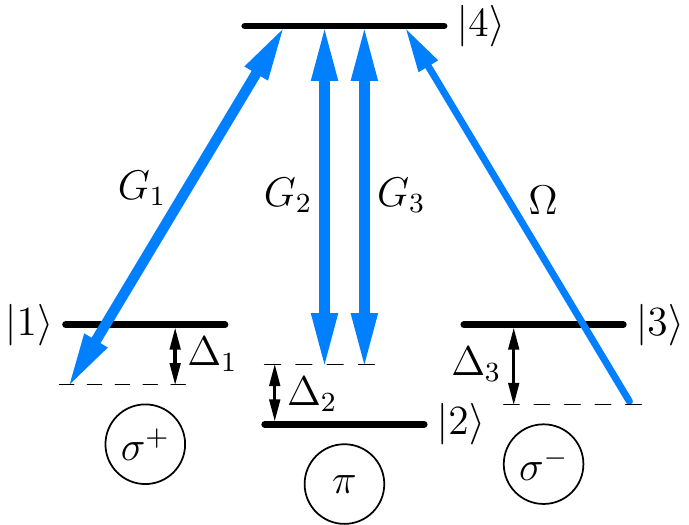}
  \caption{\label{fig:loc-scheme}
    (Color online) (a) Three standing-wave lasers $\mathbf{E}_1(z)$, $\mathbf{E}_2(x)$, and $\mathbf{E}_3(y)$, propagating in perpendicular directions, form a 3D spatial inhomogeneity, the measurement of which by means of probe laser $\mathbf{E}_4(z)$ allows one to localize the atom in an atomic state. (b) The laser polarizations and the corresponding atom-field coupling of the atomic states in a four-level tripod configuration.
  }
\end{figure}

The key role in the 3D localization scheme is played by a spatial inhomogeneity induced by three standing-wave fields $\mathbf{E}_1(z)$, $\mathbf{E}_2(x)$, and $\mathbf{E}_3(y)$, shown in Fig.~\ref{fig:loc-scheme}(a). Aligned in directions $Oz$, $Ox$, and $Oy$, respectively, these lasers invoke the Autler-Townes effect~\cite{Herkommer1997} with atomic-level splitting depending on the position of atom. The fourth probe field $\mathbf{E}_4(z)$ of a running wave propagating along $Oz$ allows one to measure the localization of an atomic level. Altogether, the lasers form the electric field
\begin{align}
  \begin{split}
    \mathbf{E}(\mathbf{r}, t) &= \mathbf{E}_1(z) e^{-i \omega_1 t} + (\mathbf{E}_2(x) + \mathbf{E}_3(y)) e^{-i \omega_2 t}
    \\
    &\quad + \mathbf{E}_4(z) e^{-i \omega_3 t} + \mbox{c.c.},
  \end{split}
\end{align}
where $\omega_m$ ($m=1,2,3$) is the laser frequency. Electric fields $\mathbf{E}_2(x)$ and $\mathbf{E}_3(y)$ are parallel to the $Oz$ axis, whereas fields $\mathbf{E}_1(z)$ and $\mathbf{E}_4(z)$ form a $\sigma^+$- and $\sigma^-$-polarized laser, respectively:
\begin{align}
  \begin{aligned}
    \mathbf{E}_1(z) &= \mathbf{e}^{-1} E_1 \sin kz, &\quad \mathbf{E}_2(x) &= \mathbf{e}^0 E_2 \sin kx,
    \\
    \mathbf{E}_3(y) &= e^{i\pi/2} \mathbf{e}^0 E_3 \sin ky, &\quad \mathbf{E}_4(z) &= \mathbf{e}^1 E_4 e^{ikz}.
  \end{aligned}
\end{align}
Here, $\mathbf{e}^q$ ($q = -1, 0, 1$) is a contravariant unitary vector, $k$ is the same wave number for each optical wave. The phase shift of $\pi/2$ was added to $\mathbf{E}_3(y)$ for convenience only.

Note that the Clebsch-Gordan coefficient~\eqref{eq:clebsch-gordan} vanishes unless the sublevels satisfy $m_F = m'_F + q$. Hence, introducing atomic levels
\begin{align}
  \begin{aligned}
    &|1\rangle = | 5^2\mathbf{S}_{1/2}, F=1, m_F=-1 \rangle,
    \\
    &|2\rangle = | 5^2\mathbf{S}_{1/2}, F=1, m_F=0 \rangle,
    \\
    &|3\rangle = | 5^2\mathbf{S}_{1/2}, F=1, m_F=1 \rangle,
    \\
    &|4\rangle = | 5^2\mathbf{P}_{3/2}, F'=0, m'_F=0 \rangle,
  \end{aligned}
\end{align}
one concludes that laser $\mathbf{E}_1(z)$ couples transition $|1\rangle \to |4\rangle$, both $\mathbf{E}_2(x)$ and $\mathbf{E}_3(y)$ couple $|2\rangle \to |4\rangle$, and $\mathbf{E}_4(z)$ couples $|3\rangle \to |4\rangle$ (see Fig.~\ref{fig:loc-scheme}(b)). We use the rotating-wave approximation (RWA) and the interaction picture, thereby the matrix elements of the atom-field coupling operator $\hbar \hat V$ are written as
\begin{align}
  \label{eq:coupling}
  \begin{aligned}
    V_{41} &= G_1 \sin kz, \quad V_{43} \equiv \Omega,
    \\
    V_{42} &= G_2 \sin kx + i G_3 \sin ky,
  \end{aligned}
\end{align}
where the Rabi frequencies
\begin{align}
  \begin{aligned}
    G_1 &= -\frac{d_{41} E_1}{\hbar}, &\quad G_2 &= -\frac{d_{42} E_2}{\hbar},
    \\
    G_3 &= -\frac{d_{42} E_3}{\hbar}, &\quad \Omega &= -\frac{d_{43} E_4}{\hbar},
  \end{aligned}
\end{align}
are given by the matrix elements of the dipole operator
\begin{align}
  d_{41} = d_{42} = d_{43} = \sqrt{\frac{1}{6}} \langle J = 1/2 \| \mathbf{d} \| J' = 3/2 \rangle.
\end{align}

If the center-of-mass position of an atom is nearly constant, then the atom can be well-localized due to the measurement of its population in an atomic level. By neglecting of the atomic kinetic energy we get the Raman-Nath approximation~\cite{Meystre1999}, and the equations of matrix-density elements are written as
\begin{subequations}
  \label{eq:dme}
  \begin{align}
    i \dot \rho_{11} &= V_{41}^* \rho_{41} - V_{41} \rho_{14} + i\gamma_1\rho_{44},
    \\
    i \dot \rho_{22} &= V_{42}^* \rho_{42} - V_{42} \rho_{24} + i\gamma_2\rho_{44},
    \\
    \label{eq:dme-rho33}
    i \dot \rho_{33} &= V_{43}^* \rho_{43} - V_{43} \rho_{34} + i\gamma_3\rho_{44},
    \\
    \begin{split}
      i \dot \rho_{44} &= V_{41} \rho_{14} - V_{41}^* \rho_{41} + V_{42} \rho_{24} - V_{42}^* \rho_{42}
      \\
      &\quad + V_{43} \rho_{34} - V_{43}^* \rho_{43} - i\gamma\rho_{44},
    \end{split}
    \\
    i \dot \rho_{12} &= V_{41}^* \rho_{42} - V_{42} \rho_{14} + (\Delta_{12} - i\Gamma_{12}) \rho_{12},
    \\
    \label{eq:dme-rho13}
    i \dot \rho_{13} &= V_{41}^* \rho_{43} - V_{43} \rho_{14} + (\Delta_{13} - i\Gamma_{13}) \rho_{13},
    \\
    \label{eq:dme-rho23}
    i \dot \rho_{23} &= V_{42}^* \rho_{43} - V_{43} \rho_{24} + (\Delta_{23} - i\Gamma_{23}) \rho_{23},
    \\
    \begin{split}
      i \dot \rho_{14} &= V_{41}^* (\rho_{44} - \rho_{11}) - V_{42}^* \rho_{12}
      \\
      &\quad - V_{43}^* \rho_{13} + (\Delta_1 - i\Gamma_{14}) \rho_{14},
    \end{split}
    \\
    \begin{split}
      i \dot \rho_{24} &= V_{42}^* (\rho_{44} - \rho_{22}) - V_{41}^* \rho_{21}
      \\
      &\quad - V_{43}^* \rho_{23} + (\Delta_2 - i\Gamma_{24}) \rho_{24},
    \end{split}
    \\
    \label{eq:dme-rho34}
    \begin{split}
      i \dot \rho_{34} &= V_{43}^* (\rho_{44} - \rho_{33}) - V_{41}^* \rho_{31}
      \\
      &\quad - V_{42}^* \rho_{32} + (\Delta_3 - i\Gamma_{34}) \rho_{34},
    \end{split}
  \end{align}
\end{subequations}
where the rest of equations are given by $\rho_{mn} = \rho_{nm}^*$, and $\sum_{n=1}^4 \rho_{nn} = 1$. Here, we introduced the frequency detunings $\Delta_m = \omega_m - \omega_{4m}$, together with their differences $\Delta_{nm} = \Delta_n - \Delta_m$. The decay rate from the upper state $\gamma = \gamma_1 + \gamma_2 + \gamma_3$, where $\gamma_m$ ($m=1,2,3$) gives the decay through channel $|4\rangle \leftrightarrow |m\rangle$. The corresponding coherent decay rates are $\Gamma_{14}$, $\Gamma_{24}$, and $\Gamma_{34}$. These rates are usually much larger than coherent decay rates between ground states, $\Gamma_{12}$, $\Gamma_{13}$, and $\Gamma_{23}$; hence, we assume that $\Gamma_{12} = \Gamma_{13} = \Gamma_{23} = 0$.

The study of the long-time limit, previously demonstrated in 1D~\cite{Paspalakis2001,Paspalakis2005} and 2D~\cite{Ivanov2010} localization schemes, makes sense for heavy atoms as the Raman-Nath regime is applied. Then, populations in atomic levels only depend on the atom-field coupling parameters, which leads to 3D localization of populations within periodic spatial domains of size $\lambda/2$, where $\lambda = 2\pi/k$ is the optical wavelength. As shown in 1D and 2D, narrow population distributions are achieved if the probe field $\Omega$ is weak.

Let us consider the localization of upper state $|4\rangle$ in the long-time limit, which in turn follows from Eqs.~\eqref{eq:dme} if $\dot \rho_{nm} = 0$. In the assumption of $\Omega = 0$, Eqs.~\eqref{eq:dme-rho13},~\eqref{eq:dme-rho23} give non-diagonal elements
\begin{align}
  \label{eq:coherences}
  \rho_{31} \approx - \frac{V_{41}}{\Delta_{13}} \rho_{34}, \quad \rho_{32} \approx - \frac{V_{42}}{\Delta_{23}} \rho_{34},
\end{align}
where correlations $\rho^*_{nm} = \rho_{mn}$ are taken into account. Because almost all the atomic population falls into ground state $|3\rangle$, one can substitute $\rho_{44} - \rho_{33} \approx -1$ into Eq.~\eqref{eq:dme-rho34} and obtain
\begin{align}
  \rho_{34} \approx \frac{V_{43}^*}{\displaystyle \frac{|V_{41}|^2}{\Delta_{13}} + \frac{|V_{42}|^2}{\Delta_{23}} + \Delta_3 - i\Gamma_{34}}.
\end{align}
Using Eq.~\eqref{eq:dme-rho33}, population $\rho_{44}$ in the second order of perturbation theory is written as
\begin{align}
  \rho_{44} \approx \frac{2}{\gamma_3} \mathop{\rm Im} \frac{|V_{43}|^2}{\displaystyle \frac{|V_{41}|^2}{\Delta_{13}} + \frac{|V_{42}|^2}{\Delta_{23}} + \Delta_3 - i\Gamma_{34}}.
\end{align}
In turn, the substitution of atom-field couplings~\eqref{eq:coupling} gives
\begin{align}
  \label{eq:upper}
  \rho_{44} \approx \frac{2\Gamma_{34} \Omega^2}{\gamma_3 (D^2 + \Gamma_{34}^2)},
\end{align}
where
\begin{align}
  \label{eq:denom}
  D = \frac{G_1^2}{\Delta_{13}} \sin^2 kz + \frac{G_2^2}{\Delta_{23}} \sin^2 kx + \frac{G_3^2}{\Delta_{23}} \sin^2 ky + \Delta_3.
\end{align}

\begin{figure}
  \includegraphics[width=7cm]{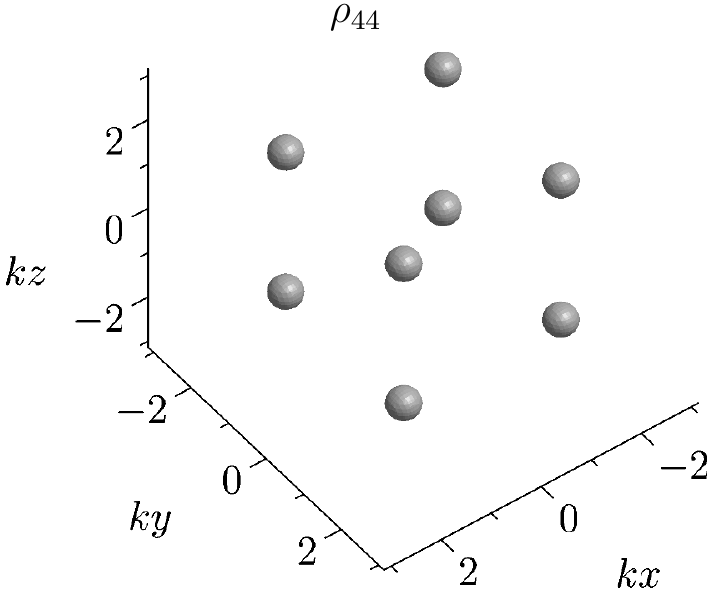}\\(a)\\
  \includegraphics[width=7cm]{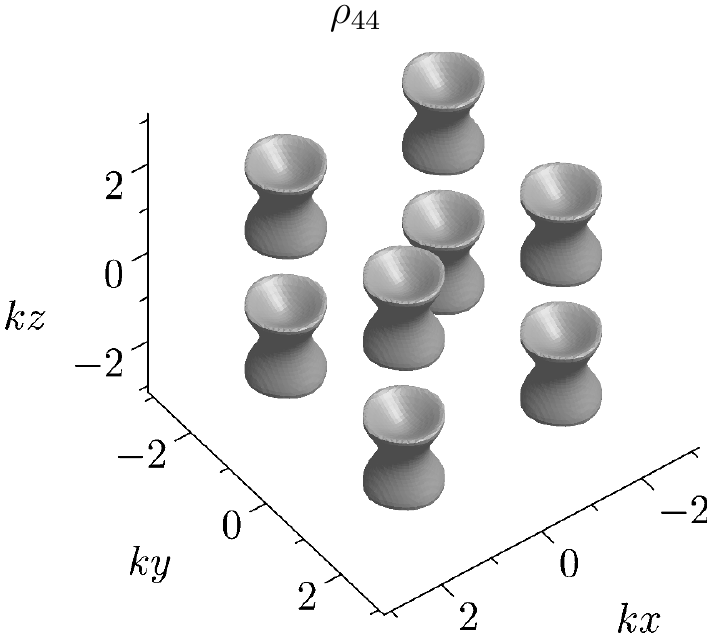}\\(b)\\
  \includegraphics[width=7cm]{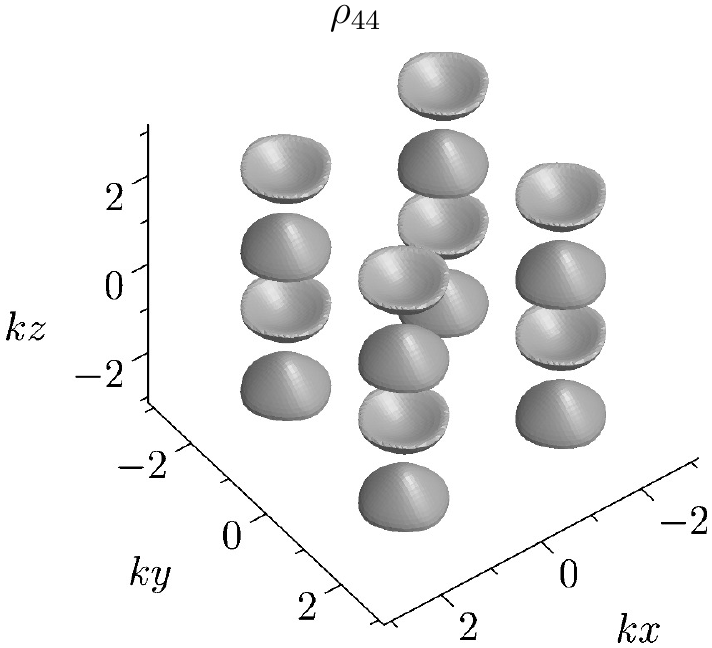}\\(c)\\
  \caption{\label{fig:main-structures}
    Isosurfaces for the upper-state population $\rho_{44} = 0.06$ as functions of $(kx,ky,kz)$. The kind of spatial structures depends on the laser frequency detunings, taking the form of (a) spheres, (b) hourglasses, (c) bowls; (a) $\Delta_1 = -3\gamma$, $\Delta_2 = -8\gamma$, $\Delta_3 = -12\gamma$; (b) $\Delta_1 = -17\gamma$, $\Delta_2 = -\gamma$, $\Delta_3 = -5\gamma$; (c) $\Delta_1 = 14\gamma$, $\Delta_2 = \gamma$, $\Delta_3 = 5\gamma$. The Rabi frequencies of standing waves are $G_1 = 6\gamma$, $G_2 = G_3 = 4\gamma$, and the Rabi frequency of probe running wave is $\Omega = 0.3\gamma$. Decay rates from the upper state are $\gamma_1 = \gamma_2 = \gamma_3 = \gamma$, coherent-decay rates through the channels $|4\rangle \leftrightarrow |m\rangle$ ($m = 1,2,3$) are $\Gamma_{14} = \Gamma_{24} = \Gamma_{34} = 1.5\gamma$.
  }
\end{figure}

In the case of $\Omega \ll G_1, G_2, G_3$, Eqs.~\eqref{eq:upper} and~\eqref{eq:denom} allow one to systematize localization structures near point $(x_0, y_0, z_0)$ which corresponds to an antinode of all strong standing waves, i.e.,
\begin{align}
  \label{eq:antinode}
  x_0 = (1 + 2l) \frac{\lambda}{4}, \quad y_0 = (1 + 2m) \frac{\lambda}{4}, \quad z_0 = (1 + 2n) \frac{\lambda}{4},
\end{align}
where $l$, $m$, $n$ are integer. Near such a point, Eq.~\eqref{eq:denom} takes the form
\begin{multline}
  D \approx - \frac{k^2 G_1^2}{\Delta_{13}} (z - z_0)^2 - \frac{k^2 G_2^2}{\Delta_{23}} (x - x_0)^2
  \\
  - \frac{k^2 G_3^2}{\Delta_{23}} (y - y_0)^2 + \Delta_3 + \frac{G_1^2}{\Delta_{13}} + \frac{G_2^2 + G_3^2}{\Delta_{23}}.
\end{multline}
Then, equality $D = 0$ gives three possible structures of the upper-state population maxima, which we adduce for
\begin{align}
  \Delta_3 + \frac{G_1^2}{\Delta_{13}} + \frac{G_2^2 + G_3^2}{\Delta_{23}} \geq 0.
\end{align}
Fig.~\ref{fig:main-structures} illustrates the following cases: (a) $\Delta_{13}, \Delta_{23} > 0$; (b) $\Delta_{13} < 0$, $\Delta_{23} > 0$; (c) $\Delta_{13} > 0$, $\Delta_{23} < 0$. Respectively, the population maxima form such spatial structures as (a) spheres, (b) hourglasses, (c) bowls. The localization distributions become narrower with an increase of $G_1$, $G_2$, and $G_3$, achieving the high-precision 3D localization in volumes substantially smaller than $\lambda^3$ when the standing-wave Rabi frequencies are large enough.

\begin{figure}
  \includegraphics[width=7cm]{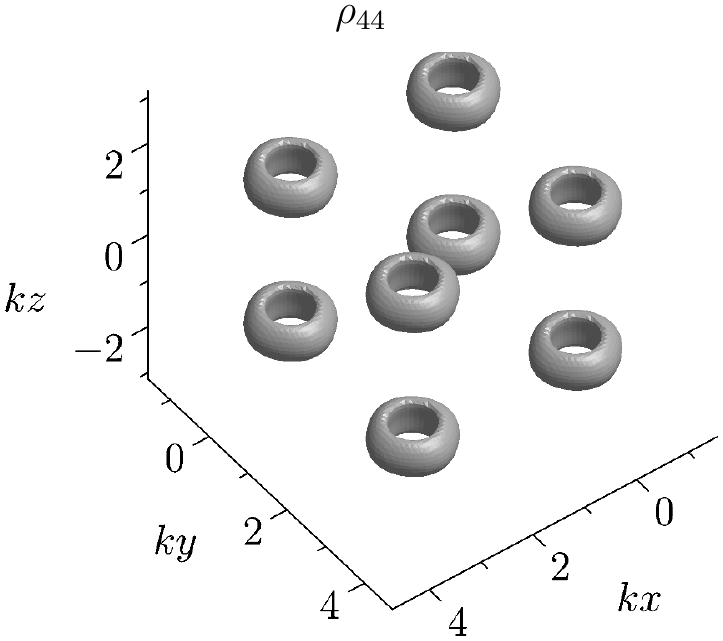}\\(a)\\
  \includegraphics[width=7cm]{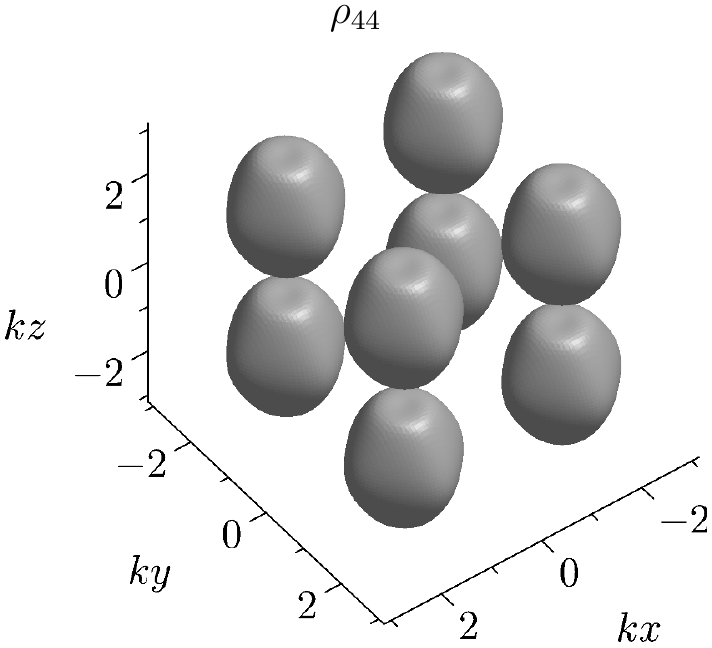}\\(b)\\
  \caption{\label{fig:structures-shifted}
    Isosurfaces for the upper-state population $\rho_{44} = 0.06$ as functions of $(kx,ky,kz)$. The kind of spatial structures depends on the laser frequency detunings, taking the form of (a) donuts, (b) deformed barrels; (a) $\Delta_1 = \gamma$, $\Delta_2 = -9\gamma$, $\Delta_3 = -5\gamma$; (b) $\Delta_1 = 10\gamma$, $\Delta_2 = -4\gamma$, $\Delta_3 = -8\gamma$. The Rabi frequencies of standing waves are $G_1 = 6\gamma$, $G_2 = G_3 = 4\gamma$, and the Rabi frequency of probe running wave is $\Omega = 0.3\gamma$. All other parameters are the same as in Fig.~\ref{fig:main-structures}.
  }
\end{figure}

In addition to localization near points~\eqref{eq:antinode}, Eqs.~\eqref{eq:upper} and~\eqref{eq:denom} demonstrate the shifting of the upper-state population by a spatial shift of $\lambda/4$. The transformation of frequency detunings $\Delta_m \to \Delta'_m$ ($m = 1,2,3$),
\begin{align}
  \label{eq:z-transform}
  \begin{aligned}
    \Delta'_1 &= 2\Delta_3 - \Delta_1 + G_1^2/\Delta_{13},
    \\
    \Delta'_2 &= \Delta_2 + G_1^2/\Delta_{13},
    \\
    \Delta'_3 &= \Delta_3 + G_1^2/\Delta_{13},
  \end{aligned}
\end{align}
reduces Eq.~\eqref{eq:denom} to the form
\begin{multline}
  D' = \frac{G_1^2}{\Delta_{13}} \sin^2\left(kz \pm \frac{\pi}{2}\right)
  \\
  + \frac{G_2^2}{\Delta_{23}} \sin^2 kx + \frac{G_3^2}{\Delta_{23}} \sin^2 ky + \Delta_3,
\end{multline}
with a shift of $\lambda/4$ along direction $Oz$ against $D$. In turn, transformation $\Delta_m \to \Delta''_m$ ($m = 1,2,3$), where
\begin{align}
  \label{eq:xy-transform}
  \begin{aligned}
    \Delta''_1 &= \Delta_1 + (G_2^2 + G_3^2)/\Delta_{23},
    \\
    \Delta''_2 &= 2\Delta_3 - \Delta_2 + (G_2^2 + G_3^2)/\Delta_{23},
    \\
    \Delta''_3 &= \Delta_3 + (G_2^2 + G_3^2)/\Delta_{23},
  \end{aligned}
\end{align}
reduces Eq.~\eqref{eq:denom} to the form
\begin{multline}
  D'' = \frac{G_1^2}{\Delta_{13}} \sin^2 kz + \frac{G_2^2}{\Delta_{23}} \sin^2\left(kx \pm \frac{\pi}{2}\right)
  \\
  + \frac{G_3^2}{\Delta_{23}} \sin^2\left(ky \pm \frac{\pi}{2}\right) + \Delta_3,
\end{multline}
shifted by $\lambda/4$ in both directions $Ox$ and $Oy$. Transformations~\eqref{eq:z-transform} and~\eqref{eq:xy-transform} lead to new localization structures; for instance, horizontally shifted spheres change to donuts, and vertically shifted hourglasses change to deformed barrels, as seen in Fig.~\ref{fig:structures-shifted}.

In conclusion, we have suggested a laser configuration consisting of three standing waves and a probe running wave, which provides the high-precision 3D localization of a four-level tripod-type atom by means of the measurement of its population. The analysis of the upper-level population shows that the corresponding 3D periodic structures depend crucially on the atom-field coupling, taking the form of spheres, hourglasses, bowls, donuts, or deformed barrels. The widths of these distributions are not limited by the optical wavelength $\lambda$, as a result, the atom can be localized in volumes substantially smaller than $\lambda^3$.

\begin{acknowledgments}
  V.~I. and Yu.~R. are grateful to Prof. Igor Sokolov (St. Petersburg State Polytechnical University) for fruitful discussions and Dr Ekaterina Efremova (St. Petersburg State University) for preliminary results of numerical simulations. This research was supported by the Government of Russian Federation, Grant 074-U01.
\end{acknowledgments}

\bibliography{localization3d}
\bibliographystyle{apsrev4-1}

\end{document}